# Electric Field Enhanced Hydrogen Storage on BN Sheet

J. Zhou[1], Q. Wang[2], Q. Sun[1,3,2,*], P. Jena[2], X.S. Chen[4]

[1] Department of Advanced Materials and Nanotechnology, Peking University, Beijing 100871, China

[2] Department of Physics, Virginia Commonwealth University, Richmond, VA 23284

[3] Center for Applied Physics and Technology, Peking University, Beijing 100871, China

[4] Shanghai Institute of Technical Physics, Chinese Academy of Science, Shanghai 200083, China

## Abstract

Using density functional theory we show that an applied electric field substantially improves the hydrogen storage properties of a BN sheet by polarizing the hydrogen molecules as well as the substrate. The adsorption energy of a *single* $H_2$ molecule in the presence of an electric field of 0.05 a.u. is 0.48 eV compared to 0.07 eV in its absence. When one layer of $H_2$ molecules is adsorbed, the binding energy per $H_2$ molecule increases from 0.03 eV in the field-free case to 0.14 eV/$H_2$ in the presence of an electric field of 0.045 a.u. The corresponding gravimetric density of 7.5 wt % is consistent with the 6 wt % system target set by DOE for 2010. Once the applied electric field is removed, the stored $H_2$ molecules can be easily released, thus making the storage reversible.

---

Hydrogen, the simplest and the most abundant element in the universe, is an energy carrier and is expected to play a critical role in a new, decentralized energy infrastructure with many important advantages over other fuels. Unlike fossil fuels such as oil, natural gas and coal that contain carbon, produce $CO_2$, contribute to global warming and have limited supply, hydrogen is clean, abundant, non-toxic, renewable and packs more energy per unit mass than any other fuel. However, its commercial use as an alternate energy has substantial difficulties to overcome. Among these, the most difficult challenge is to find materials that can store hydrogen with large gravimetric and volumetric density and operate under ambient thermodynamic conditions. The Department of Energy's system target for the ideal hydrogen storage material is that the gravimetric density of hydrogen should reach 6 wt % by 2010. In addition, the storage materials should be able to reversibly adsorb/desorb $H_2$ in the temperature range of -20 to 50 °C and under moderate pressures (max. 100 atm). The first requirement limits the choice of storage materials to be composed of elements lighter than Al, while the later requires hydrogen binding energies to be between physisorption and chemisorption energies. Recently Bhatia and Myers [1] have studied the optimum thermodynamic conditions for hydrogen adsorption by employing the Langmuir equation and derived relationships between the operating pressure of a storage tank and the enthalpy of adsorption required for storage near room temperature. They have found that the average optimal adsorption enthalpy should be 15.1 kJ /mol, if operated between 1.5 and 30 bar at 298 K. When the pressure is increased to 100 bar, the optimal value becomes 13.6 kJ/mol. Therefore the optimal adsorption energy for $H_2$ should be in the range of 0.1 ~ 0.2 eV/$H_2$. Unfortunately, the above two requirements are difficult to satisfy simultaneously. The bonding of hydrogen in light elements is either too strong, as in light metal hydrides and organic molecules, or too weak, as in graphite and carbon and BN fullerenes and nanotubes [2-5]. For example, although $BH_3$, $NH_3$, and $B_{12}H_{12}$ contain high weight percentage of hydrogen, the kinetics of hydrogen release is poor due to strong H bond.

To overcome the above problems and to balance both energetics and weight percentage, recent efforts have been devoted to design new materials with exposed metal sites following mechanisms proposed by Kubas [6] and Jena and coworkers [7]. The Kubas mechanism [6] takes advantage of the unfilled d-orbitals of transition metal atoms where $H_2$ molecules donate electrons to the unfilled d-orbitals and the transition metals back donate the electrons to the $H_2$ molecules. As a result $H_2$ retains its molecular bond, but in a slightly stretched form. The work of Jena and coworkers [7], on the other hand, focused on metal cations where stretching of the molecular bond of $H_2$ is caused by charge polarization. The difficulty with the former approach is that transition metal atoms tend to cluster [8-11] while in the later approach the binding energy tends to be lower than the desired value especially when hydrogen coverage increases [12-13]. Recent attempts have concentrated in using metal ions that carry multiple charges such as $Mg^{2+}$ and $Al^{3+}$ in light porous materials [14-16]. Three strategies have aimed at introducing the exposed metal sites [14] in one of three ways: (1) Removing the metal-bound volatile species which typically function as terminal ligands; (2) Incorporating metal species; and (3) Impregnating materials with excess metal cations. However, experimentally these techniques are complicated and difficult to control as any other available atoms or molecules in the environment would saturate the exposed metal site, thus diminishing expected hydrogen adsorption.

In this Letter we propose a different approach that would make materials synthesis less complicated while improving the thermodynamics and reversibility of hydrogen storage. Note that the mechanism that allows exposed metal cations to store hydrogen in quasi-molecular form is the polarization of the $H_2$ molecule caused by the electric field associated with point ions [7,11]. Can the same objective be achieved by directly applying an external electric field to hydrogen storage materials? We show that it is indeed possible by considering a single BN sheet which has a hexagonal graphene-like structure. Using first-principles calculations we find that in an applied electric field hydrogen molecules, behaving like electric dipoles, as well as electrons in BN sheet are polarized. The electrostatic interactions between hydrogen molecules and BN sheet greatly improve the storage performance. For example, in an electric field of 0.05 a.u. the binding energy of $H_2$ is 0.48 eV/$H_2$ when a *single* $H_2$ molecule is introduced. When one layer of $H_2$ is adsorbed on the BN sheet, although the binding energy decreases to ~ 0.14 eV/$H_2$, it is five times larger than the binding energy of 0.03 eV/$H_2$ in the absence of the E-field. The corresponding gravimetric density of hydrogen is 7.5 wt %. Once the applied electric field is removed, the system goes back to its original state and hydrogen molecules easily desorb, yielding good reversibility and fast kinetics.

The reasons for choosing BN sheet for our study are the following: Since the discovery that graphene, a single sheet of graphite, can be experimentally isolated [17] considerable attention has been focused on synthesizing similar BN-based materials because of their structural similarities with carbon. Although *h*-BN manifests a structural variety similar to that of *h*-C and possesses the same crystal structure with very close cell parameters, their electronic properties are distinctly different. For example, *h*-C can be metallic, semiconducting, or semimetallic depending upon their dimensionality, size, and coordination. *h*-BN, on the other hand, is typically a wide-band gap semiconductor. Analogous to carbon atoms in graphite, boron and nitrogen atoms in *h*-BN form a 2D honeycomb structure with strong covalent bonds in the plane and weak bonds between different planes. Recently there is considerable interest in synthesizing BN layered structures. For example, well ordered *h*-BN layers can be grown by the thermal decomposition of borazine ($B_3N_3H_6$) on the surface of a transition metal, such as lattice-matched Ni (111) or lattice-mismatched Rh (111) [18-20]. The layers interact weakly with the metal, but they are stable to high temperatures (up to 1000 K) in air. When using a micromechanical cleavage method [21-22], a 2D *h*-BN with few atomic layers (more than five layers) can be obtained. Very recently Han and coworkers [23] successfully synthesized a *mono-atomic-layered* B-N sheet by using a chemical-solution-derived method starting from single-crystalline hexagonal boron nitride. Currently the single-layered BN sheet has become a new nano-platform for manipulating the electronic structures and magnetism for various applications [24-27].

Our calculations are based on density functional theory (DFT) with generalized gradient approximation (GGA) for exchange and correlation potential for which Becke-Lee-Yang-Parr (BLYP) [28] functional is used as implemented in DMol3 package [29-31]. The Dmol3 code has been extended to include not only the potential arising from the nuclear charges as "external" potential, but also the static potentials arising from an externally applied electric field. The Hamiltonian is composed of the potential arising from the external electric field ($E_{ext.}$), kinetic operator $T$, and Hatree ($V_H$), and exchange-correlation ($V_{xc}$) potential.

$$H = T + V_H + V_{xc} - eE_{ext} \tag{1}$$

Using this prescription Delley [32] has studied the dissociation of molecules in strong electric fields and found that the bond length and vibrational frequency as a function of the field can be fit very well all the way up to the dissociation limit by an analytical formula derived from a Morse potential model including an additional external electric field term. The DMol3 package is widely used in molecular systems to study the effects of electric field on reactivity [33], hyperpolarizability [34], work function [35], and field-emission [36]. Furthermore, the Hamiltonian in Eq. (1) can be extended to a periodic system [30, 31] in a straightforward manner for studying the effects of a uniform constant external electric field on conductance [37], geometry [38], energy barrier [39], electron-trapping [40], and energy band-gap [41].

All our calculations have been carried out using periodic boundary condition to simulate an infinite BN sheet. The vacuum space of 15 Å is used in the direction normal to the BN sheet in order to avoid interactions between two layers. Integrations over reciprocal space are based on Monkhorst-Pack scheme with 2x2x1 and 9x9x1 k-points grid meshes when either a single or a layer of $H_2$ is considered, respectively. We used semi-core pseudopotential with double-numerical basis set plus d functions (DND). The orbital cutoff was set to be global and with a value of 3.4 Å. The atoms were relaxed without any symmetry constraints. Convergence in energy, force, and displacement was set at $2*10^{-5}$ Ha, 0.001 Ha/Å and 0.005 Å, respectively. The accuracy of our exchange-correlation functional and basis set was tested through the following data. The bond length, binding energy and polarizability of the $H_2$ molecule parallel to its axis are calculated to be 0.741 Å, 4.42 eV, and 6.0 a.u, respectively which agree well with corresponding experimental values of 0.741 Å, 4.45 eV, and 6.3 a.u. The in-plane lattice parameter of BN sheet has been optimized and the calculated value of 2.5 Å is also in good agreement with the experimental value of 2.505 Å.

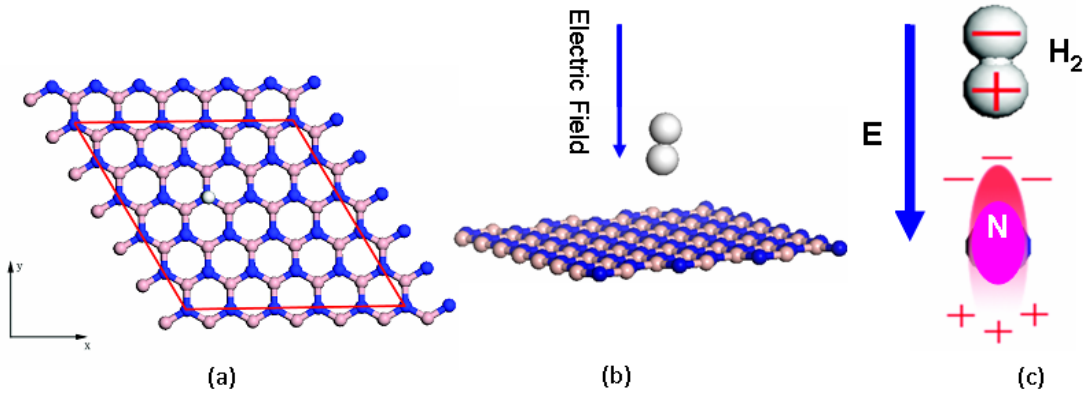

FIG. 1. A single $H_2$ molecule adsorbed on the BN sheet: (a) the top view where the red lines show the supercell used in the simulations, (b) the side view, where vertical E-field is applied in the -z direction, (c) the model demonstrating the polarized $H_2$ molecule and the polarized electrons at N site on the BN sheet.

We first consider the interaction between a *single* $H_2$ molecule and the BN sheet. The relative stability of different configurations of $H_2$ molecule on the BN sheet in an external electric field (selected as 0.040 a.u.) vertical to the sheet has been studied (Fig. 1). We tested four adsorption sites of $H_2$ molecule on the BN sheet: directly on top of a B or N atom (on-top site), above middle of a bond linking B-N atoms (bridge site), and above the center of a honeycomb-like hexagon (hollow site). For each site, various initial directions of the $H_2$ molecule have been studied. After optimization, we found that $H_2$ molecule prefers to align along the z direction, parallel to the E-field. The lowest energy is found when the $H_2$ molecule adsorbs on the N atom. The reason for this will be discussed in the following. We also checked the cases when electric field is applied along x and y directions, but found no obvious improvement for H adsorption. So in the following discussions, we focus only on the E-field vertical to the BN sheet.

With the help of Mulliken population analysis, we identified that the upper H atom is negatively charged, while the charge on the lower one is positive. This can be understood by the polarization effects of the E-field, since it "pushes" positive charges along its direction. Due to its large electronegativity, N attracts electrons from B. Thus, on the BN sheet the N sites are electron-rich and are easily polarized in the applied electric field in a way similar to the $H_2$ molecule. The electrostatic interactions between the $H_2$ and N make N the favored site for $H_2$ absorption (Fig. 1c)

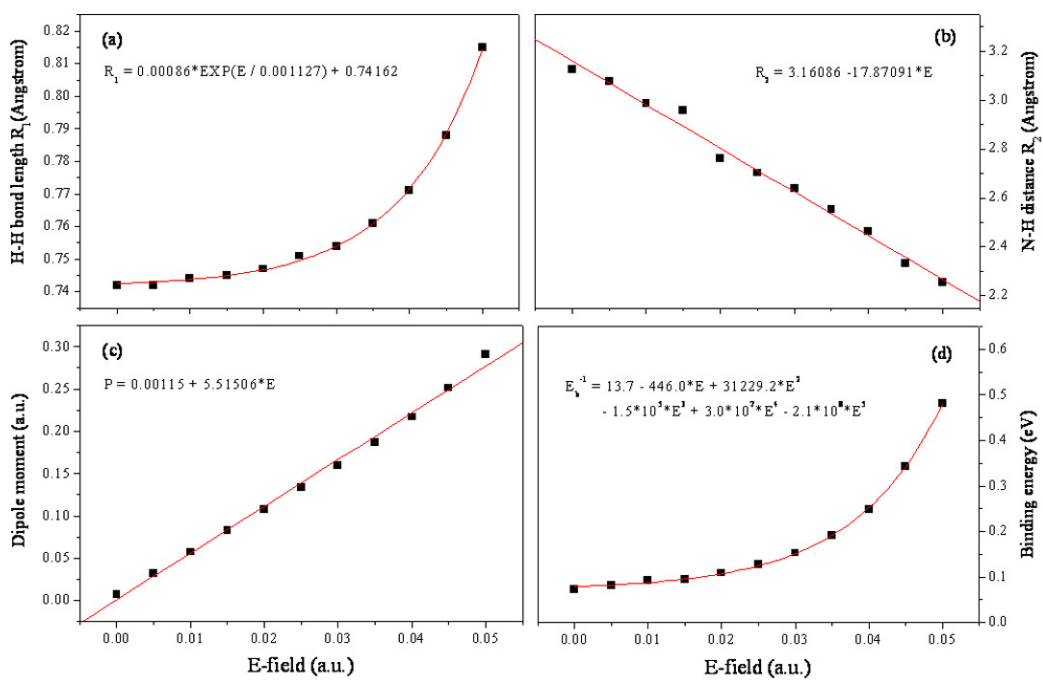

FIG. 2. Calculated (black points) and fitted curve (red) of (a) H-H bond length $R_1$ (Å), (b) N-H distance $R_2$ (Å), (c) the induced dipole moment of $H_2$ molecule, and (d) binding energy as a function of electric field strength.

We have further studied the effect of electric field strength on hydrogen adsorption by focusing on the N "on top-site". The H-H bond length ($R_1$), distance between N atom and $H_2$ molecule ($R_2$), the dipole moment of $H_2$ molecule, and binding energy per $H_2$ are plotted as a function of the magnitude of the E-field in Figure 2. The data are fitted with the red curves. We see that the H-H bond length increases exponentially with increasing electric field. This is due to the polarization interaction induced by the electric field and concurrent binding with the BN sheet. Simultaneously, the distance between the $H_2$ molecule and the substrate decreases linearly as the E-field increases. This indicates that the interactions between $H_2$ molecule and the BN sheet can be easily tuned by the E-field. When the E-field reaches 0.050 a.u., the H-H bond length $R_1$ becomes 0.815 Å and the distance to the substrate is 2.255 Å, resulting in adsorption energy of 0.48 eV. When the electric field reaches 0.060 a.u., the $H_2$ molecule is dissociated and atomically bonded to the BN sheet, indicating that molecular hydrogen adsorption can be achieved only below a certain critical electric field. Furthermore, from Fig. 2c we find that the induced dipole moment of $H_2$ molecule changes linearly with E-filed. Thus, in a weak electric field, contributions from higher order of nonlinear terms can be neglected.

We also analyzed the adsorption energy, defined as the difference between energies of the complex and the separated ones in the external electric field, i.e., $E_b = E(BN) + E(H_2) - E(BN+H_2)$ quantitatively. We found that the adsorption energy changes with the applied field in a complicated way (Fig. 2d):

$$1/E_b = 13.7 - 446.0 * E + 31229.2 * E^2 - 1.5 * 10^5 * E^3 + 3.0 * 10^7 * E^4 - 2.1 * 10^8 * E^5$$

According to classical physics, due to the dipole-dipole interaction, E scales with the distance R as $E \sim 1/R^3$. Because of the complicated induced polarization, in our case the relationship scales as $E \sim 1/R^{5.5}$ as shown in Fig. 3.

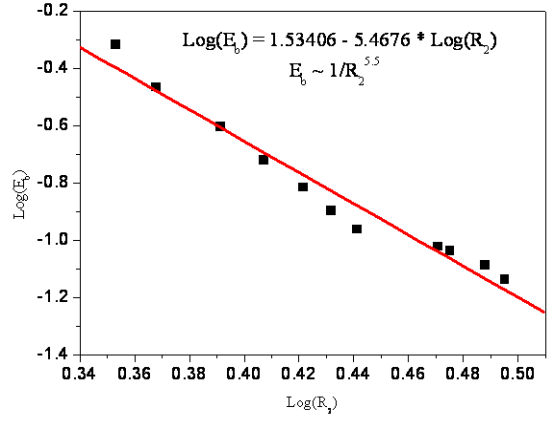

Fig. 3 Changes of adsorption energy $E_b$ with distance $R_2$.

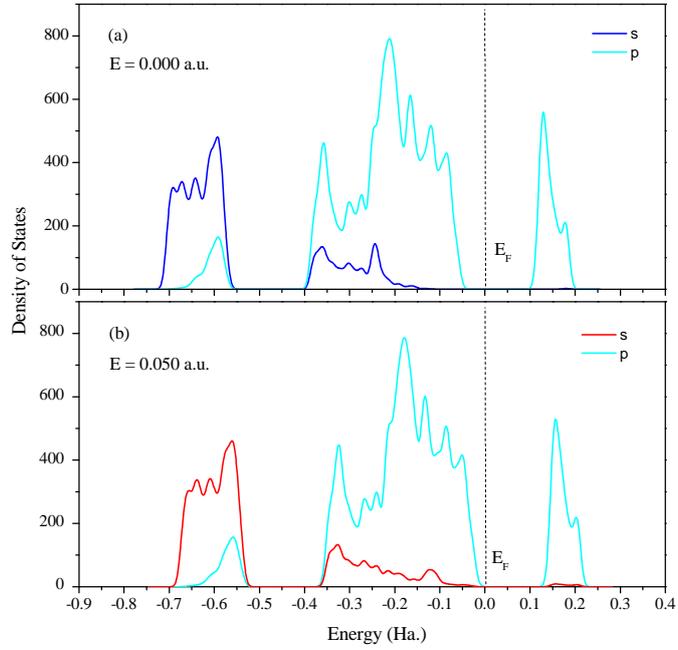

FIG. 4. PDOS of BN sheet and $H_2$ molecule are displayed with and without E-field. The Fermi level is denoted as the vertical dashed line.

To investigate the effect of the electric field on the electronic structure, we plot the partial density of states (PDOS) of the system with and without the E-field (Fig. 4). In the absence of the electric field (Fig. 4a), the semiconductor nature of the BN sheet can be clearly seen. The corresponding energy band gap of 3.52eV is less than that of the gap in the bulk phase, namely 5.97eV [42]. When the E-field of 0.050 a.u. is applied, the adsorption energy of $H_2$ molecule is increased to 0.48 eV, but the semiconductor property of the system is still retained (Fig. 4b), indicating that the BN sheet is not electrically broken down in such an applied electric field. We also see that the peak in the s density of states at ~ -0.23 a.u. contributed by the $H_2$ s orbital in the absence of the field shifts to higher energy range (~ -0.12 a.u.) when the E field is applied. Only 2p orbitals of B and N atoms contribute to DOS at the Fermi level. The iso-surface of the electrostatic potential in the E-field of 0.05 a.u. is shown in Fig. 5.

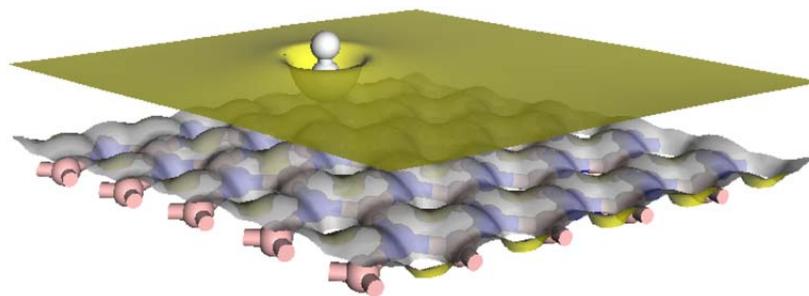

FIG. 5. Iso-surface of the electrostatic potential of one $H_2$ molecule absorbed on BN sheet in E-field of 0.05 a.u.

Next we study the case when one layer of $H_2$ molecule is adsorbed on the BN sheet (Fig. 6). We initially placed $H_2$ molecules with various orientations on four different adsorption sites as mentioned above. Once again, we found that the optimized $H_2$ molecules remain vertical to the sheet, i.e. parallel to the external electric field and the N "on top" sites remain as the favored adsorption sites. The changes of binding energy per $H_2$ molecule with different E-field are shown in Table 1 and Fig. 6c. We see that the binding energy increases from $0.03 eV/H_2$ in the absence of the E-field to $0.14 eV/H_2$ with a field of 0.045 a.u. This adsorption energy is already within the energy window required for applications under ambient thermodynamic conditions [1]. Equally important, the gravimetric density of stored hydrogen is 7.5 wt %. This is consistent with the 6 wt % system target set by DOE for 2010. To check the reversibility of hydrogen storage, we removed the external electric field and, as expected, found that all the adsorbed $H_2$ molecules become weakly bound with a binding energy of 0.03 $eV/H_2$. Therefore, storage and release of hydrogen can be easily achieved via switching on /off the external electric field.

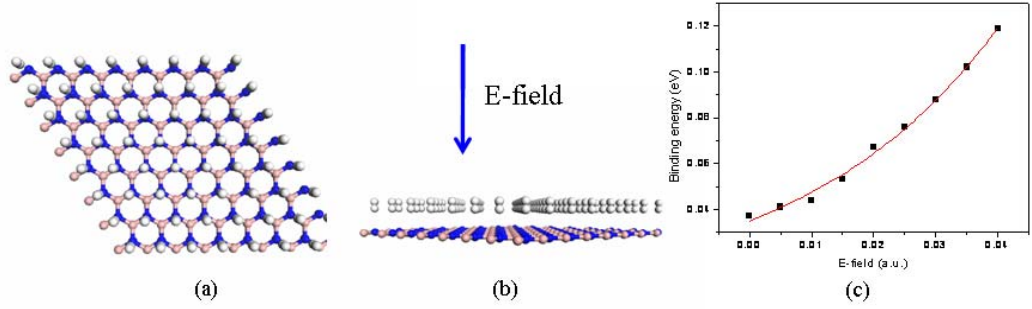

FIG. 6. A layer of $H_2$ molecules adsorbed on the BN sheet: (a) the top view, (b) the side view, (c) the changes of binding energy with E-field.

Table 1. Binding energy (eV) per $H_2$ molecule under various electric fields (a.u).

| E-field | 0.000 | 0.005 | 0.010 | 0.015 | 0.020 | 0.025 | 0.030 | 0.035 | 0.040 | 0.045 |
|---|---|---|---|---|---|---|---|---|---|---|
| $E_b$ | 0.037 | 0.041 | 0.044 | 0.053 | 0.067 | 0.076 | 0.088 | 0.102 | 0.119 | 0.140 |

To assess the feasibility that our study can lead to the focused discovery of new hydrogen storage materials for practical applications, it is necessary to examine some of the assumptions we have made. First, we have chosen a *single* BN sheet while a real material may contain several layers. As we have pointed out in the beginning, a *single* atomic BN layer has been synthesized experimentally [23]. In fact, before the experimental synthesis, a single atomic BN layer was used as a theoretical model for hydrogen storage [43]. However, it was found that the adsorption energy is only 0.04eV/$H_2$, in agreement with our result of 0.03eV/$H_2$ in field-free case. Here we show that the applied external electric field can significantly enhance the hydrogen adsorption on BN sheet. Since the interaction between the different BN layers is weak, we do not expect that the characteristics of hydrogen adsorption induced by an externally applied electric field will change. Nevertheless, we are currently studying this system with multiple layers. One may also wonder if graphene can be used to store hydrogen under the influence of an electric field. We indeed first began our study by using graphene but could not succeed due to poor convergence. The reason for our success with the BN sheet probably lies with its slightly ionic character where the electronegative N atom draws electrons and hence facilitates polarization under an applied electric field. Second, we have used density functional theory that does not treat long range interactions accurately [44] and hence the accuracy of the calculated energetics of weakly bound systems can be called into question. Recently several authors [45, 46] have incorporated long range interactions into DFT to treat weakly bound systems. They have found that the inclusion of long range interaction *increases* the binding energies obtained using GGA functionals in DFT. This means that the current calculated binding energies can, at the worst, be an underestimate. Since the computed binding energies are 0.14 eV/$H_2$ and for applications the desirable range is around 0.2 eV/$H_2$, we feel confident that application of an electric field to store hydrogen in a polarizable medium such as BN sheet has merit.

In conclusion, we have shown that application of an external electric field provides a new path way to store hydrogen which has the following advantages: (1) Contrary to the local electric field produced by exposed metal sites embedded in a substrate or a matrix, external applied electric field is more effective and provides better control for reversible hydrogen storage. In particular, the energetics and kinetics can be manipulated by tuning the field strength. (2) This process avoids complicated synthesis routes for embedding metal ions and does not suffer from the possibility that these ions may either cluster during repeated hydrogenation/dehydrogenation process or be poisoned by other gases such as oxygen. (3) In contrast to the ionic hydrides and alanates where the release of $H_2$ might render the materials unstable or incapable of re-absorption of $H_2$ if its chemical nature is altered, here the $H_2$ release can be very easily achieved by removing the electric field. (4) The reversibility, kinetics, and material stability can be controlled by choosing appropriate electric field strength. We hope that our finding will stimulate new experimental work in this direction.

## Acknowledgment

This work is partially supported by grants from the National Natural Science Foundation of China (NSFC-10744006, NSFC-10874007) and by the US Department of Energy. The authors thank B. Delley for discussions.